# Silicon-based photonic crystal nanocavity light emitters

Maria Makarova, Jelena Vuckovic, Hiroyuki Sanda, Yoshio Nishi
*Department of Electrical Engineering, Stanford University, Stanford, CA 94305-4088*

**Abstract**

We have demonstrated an up to seven-fold enhancement of photoluminescence from silicon-rich silicon nitride film due to a single photonic crystal cavity. The enhancement is partially attributed to the Purcell effect. Purcell factor predicted by FDTD calculations is 32 for a linear three-defect cavity mode with computed quality factor of 332 and mode volume of 0.785 $(\lambda/n)^3$. Experimentally measured cavity quality (Q) factors vary in the range of 200 to 300, showing excellent agreement with calculations. The emission peak can be tuned to any wavelength in the 600 to 800 nm range.

Silicon-based light sources compatible with the mainstream CMOS technology are highly desirable because they will have a low manufacturing cost relative to III/V semiconductor diodes, and it will be easier to integrate them with electronic components on the same chip. Photoluminescence (PL) from silicon-rich silicon nitride and, more commonly, oxide films with 3 to 5 nm precipitates of silicon nano-crystals (Si-nc) in the dielectric matrix, has been studied.[1,2] The luminescence is attributed to confined exciton recombination in the Si-nc or to the radiative recombination centers located at the interface between the Si-nc and the dielectric.[2] Internal quantum efficiencies can be as high as 59%,[3] and optical gain for Si-ncs has been demonstrated.[2] Confining luminescent material in an optical micro-cavity enhances the emission by restricting the resonant wavelength to a directed radiation pattern that can be collected effectively, and by reducing radiative lifetime of the on-resonance emitters due to the Purcell effect.[4,5] Reduction in radiative lifetime is particularly important for the development of lasers based on Si-ncs because it makes radiative recombination compete more favorably with non-radiative recombination processes which increase at higher pump powers. In this paper, we demonstrate light emitters based on two-dimensional (2-D) photonic crystal (PC) cavities fabricated in silicon-rich silicon nitride membrane. We used 2-D PC nanocavities because of their high Q-factor (Q) values and small mode volumes (V), since both are necessary for the Purcell effect. Planar geometry of the implementation is well suited for integration with other optical devices on a chip.

The structures were fabricated starting from bare silicon wafers. At the first step, a 500-nm-thick oxide layer was formed by wet oxidation. At the second step, a 250-nm-thick layer of silicon-rich silicon nitride was deposited by a chemical vapor deposition from $NH_3$ and $SiH_2Cl_2$ gases at 850°C. Next, a positive electron beam resist, ZEP, was spun on a wafer piece to form a 380-nm-thick mask layer. Photonic crystal pattern was exposed on the Raith 150 electron beam system. After development, the pattern formed in the resist layer was transferred into the silicon nitride layer by reactive ion etching with $NF_3$ plasma[6] using ZEP pattern as a mask. All remaining resist was removed by oxygen plasma. The oxide layer was removed under photonic crystal structures by the 6:1 buffered oxide etch. Fabricated PC cavity membrane with periodicity (**a**) of 330 nm is shown in the insert on Fig. 1.

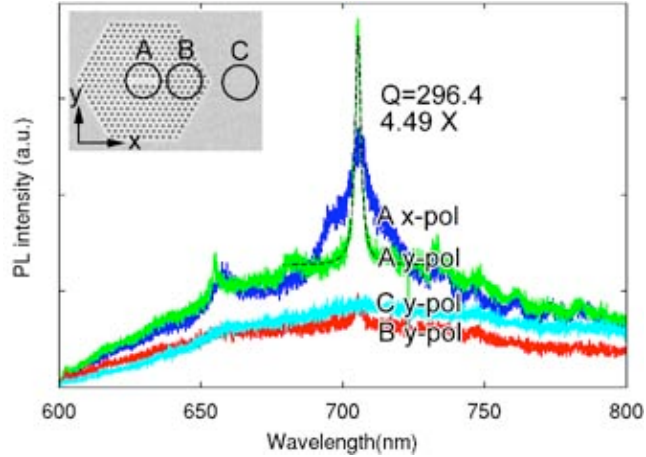

FIG. 1. Polarized PL spectra from the areas shown in the insert: cavity region (A), PC region (B), and unpatterned film (C). Dashed line shows Lorentzian fit to y-polarized cavity resonance with Q=296.4. The emission with y-polarization from region A is enhanced 4.49 times relative to region C at resonant wavelength of 705.2 nm.

Micro-photoluminescence setup was used to measure radiation spectra from the fabricated structures. A single 100X objective lens with NA=0.5 was used to image the sample with white light for alignment, to focus the pump beam, and to collect luminescence in the vertical direction (perpendicular to photonic crystal membrane). A 5-mW 532-nm green laser was used as the excitation source. The beam was spatially filtered though a pinhole to achieve the small spot diameter of about 1μm necessary for selective excitation on the sample. The total incident pump power was about 0.3 mW.

Polarized PL spectra from a single PC cavity structure with periodicity of 330 nm are shown in Fig. 1. The spectra were taken from three locations on the structure marked by circles in the insert on Fig. 1 by selectively exciting the regions, and spatially filtering the signal so that only PL coming from the region of interest was detected. For the cavity region (A) emission with polarizations along the cavity length (x-pol) and perpendicular to it (y-pol) were measured. For the PC region (B) and unpatented region (C) only the y-polarization is plotted for comparison with the stronger y-polarized resonance of the cavity. The intensity at the resonant wavelength of 705.2 nm is increased 4.5 times relative to that of the unpatterned film for the electric field polarized along y-direction. Seven-fold intensity enhancement was observed without polarization selection. Lorentzian fit, shown as a dashed line, gave Q=296 for the y-polarization.

A number of cavities of the same design were fabricated on the same chip. The measured quality factors fell in the range from 200 to 306. To tune the resonance location, structures with slightly different hole radii were produced by varying electron beam exposure dose. The resonance wavelength shifted from 680 to 720 nm as the hole radius changed from 132 to 122 nm.

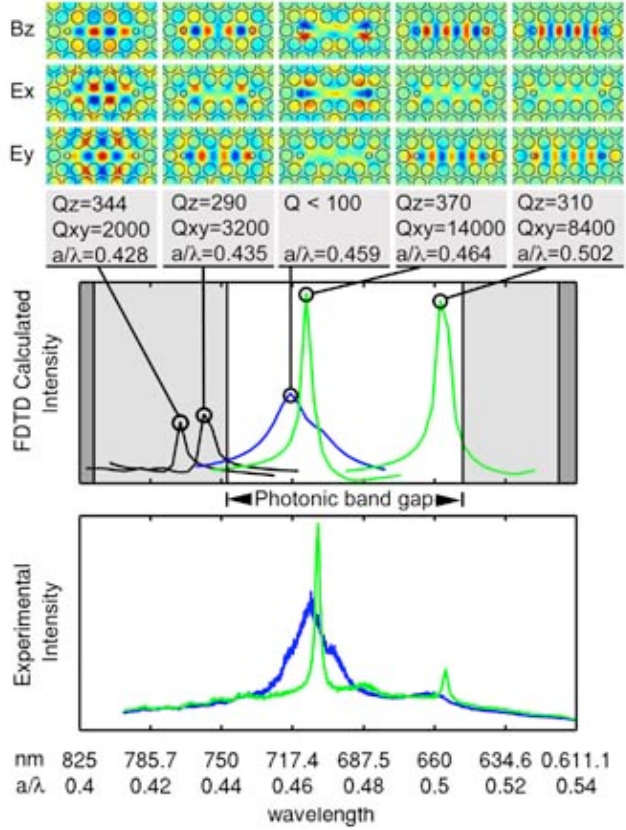

FIG. 2. Electromagnetic field distributions for TE cavity modes and their Q-factors in-plane (Qxy) and out-of-plane (Qz) calculated by FDTD together with calculated and measured spectra plotted on the same wavelength scale. The axes are: z -- perpendicular to the membrane, x -- along the cavity axis, and y -- perpendicular to x and to z. Photonic band gap for TE modes is indicated (white region). The region shaded in light gray indicates the span of the photonic band edges in frequency from X point to J point.

The optical properties of the photonic crystal cavities were analyzed using three-dimensional finite difference time-domain (3D FDTD) calculation method.[7] The modeling parameters were chosen to closely resemble fabricated structures with refractive index of 2.11, as measured by spectroscopic ellipsometry at 700 nm, photonic crystal slab thickness of 0.75**a**, and hole radius of 0.4**a**. Fig. 2 shows the electromagnetic field distributions for TE cavity modes and their in-plane (Qxy) and out-of-plane (Qz) Q-factors calculated by FDTD, and both, calculated and measured spectra, which are plotted in the same wavelength scale. The axes are: z -- perpendicular to the membrane, x -- along the cavity axis, and y -- perpendicular to x and to z. PC exhibits a 19% band gap for TE modes, from 0.4416 to 0.535 $a/\lambda$, as indicated on the Fig. 2. Generally, the high Q mode observed for the three-hole defect PC cavities in high refractive index materials has four lobes of magnetic field[8]. Here, this mode is at 0.428 $a/\lambda$ and is outside the complete photonic band gap as it falls below the band edge at J-point. The next order mode with five lobes of magnetic field is also below the band edge. There are only slight hints of these modes in the measured spectra. Experimentally observed frequency and polarization for the three modes that fall into the complete photonic band gap are in excellent agreement with theoretical calculations. The broad mode at 0.459 $a/\lambda$ is primarily polarized in the x-direction as evident from its electric field distribution, and matched by experimental measurement. The next two modes at 0.464 a/λ and 0.502 a/λ are polarized along the y-axis according to their electric field distribution and are observed with this polarization experimentally at 0.467 a/λ and 0.503 a/λ respectively. The slight discrepancy in frequency may be attributed to the slight deviation between the fabricated structure and the model. The most prominent mode in the measured spectrum is the highest Q mode at 0.467 a/λ which has six lobes of Bz in the cavity, calculated mode volume of 0.785 $(\lambda/n)^3$, calculated Q of 332, and maximum radiative rate enhancement of 32, as given by the Purcell factor, $F=3/(4\pi^2) (\lambda/n)^3 Q/V$[4,5]. Using experimentally measured Q of 296 the Purcell factor becomes 28.5. The actual observed enhancement depends on how many emitters fall into the electric field maxima and within the cavity resonance, on how well their dipole moments are aligned with the electric field, and on what is the collection efficiency of the cavity mode relative to unpatented film. The experimentally observed PL enhancement is 4.5 times at the resonant wavelength, a factor of 6.3 lower than the theoretical maximum possible. This is expected, because the experimentally observed value is an averaged value of the Purcell factor for all emitters, and majority of them are spectrally and spatially detuned from the cavity resonance, and therefore do not exhibit a maximum Purcell factor.

In summary, we have demonstrated an enhancement of PL from silicon-rich silicon nitride film with a single PC cavity for the first time, to the best of our knowledge. The use of silicon nitride film rather then more commonly employed silicon oxide film with Si-ncs allows higher index contrast necessary for stronger optical confinement in PC cavities. Studied cavities show excellent agreement with theory. The observed PL enhancement is especially important because it results from the strong Purcell effect and thus can shorten the radiative lifetime of emitters considerably, which could be crucial for making a laser based on Si-ncs. Theoretically much higher Q-factor cavities can be realized in the material system reported here, so even stronger enhancement of PL can be achieved. We would like to emphasize that the processing used to fabricate these light sources is fully compatible with CMOS fabrication technology, so optical and electronic components could be seamlessly integrated on a single chip at low cost. This may open the door to a variety of applications ranging from optoelectronics to biophotonics, especially since the emission wavelength can be chosen anywhere from around 600 to 850 nm.

This work has been supported in part by the CIS Seed Fund, MARCO Interconnect Focus Center, and DARPA nanophotonics seed fund.

**References:**

[1] L. Dal Negro, J. H. Yi, L. C. Kimerling, S Hamel, A. Williamson, and G. Galli, Appl. Phys. Lett. **88**, 183103 (2006)

[2] L. Pavesi, Proc. of SPIE **6125**, 612508, (2006)

[3] R. J. Walters, J. Kalkman, A. Polman, H. A. Atwater, and M. J. A. de Dood, Phys. Rev. B **73**, 132302 (2006)

[4] E. M. Purcell, Phys. Rev. **69,** 681 (1946).

[5] J. Vuckovic, C. Santori, D. Fattal, M. Pelton, G.S. Solomon, and Y. Yamamoto, "Cavity enhanced single photons from a quantum dot," book chapter in *Optical microcavities*, edited by K. Vahala, (World Scientific, 2004)

[6] O. Levi, W. Suh, M. M. Lee, J. Zhang, S. R. J. Brueck, S. Fan, and J. S. Harris, Proc. SPIE **6095**, 6095-24 (2006).

[7] J. Vuckovic, M. Loncar, H. Mabuchi, A. Scherer, Phys. Rev. E **65**, 016608, (2002)

[8] Y. Akahane, T. Asano, B. Song, S. Noda, Nature **425**, 944 (2003)